\documentclass[]{pasj01}

\begin{document} 
\Received{2018/03/03}
\Accepted{2018/07/17}

\title{A 100-pc Scale, Fast and Dense Outflow in Narrow-Line Seyfert 1 Galaxy IRAS~04576+0912}

\author{Toshihiro \textsc{Kawaguchi}\altaffilmark{1}%
}
\altaffiltext{1}{Department of Economics, Management and Information Science, 
Onomichi City University, 
Hisayamada 1600-2, Onomichi, Hiroshima 722-8506, Japan}
\email{kawaguchi@onomichi-u.ac.jp}

\author{Shinobu \textsc{Ozaki}\altaffilmark{2}}
\altaffiltext{2}{National Astronomical Observatory of Japan, Mitaka, Tokyo 181-8588, Japan}

\author{Hajime \textsc{Sugai}\altaffilmark{3}}
\altaffiltext{3}{Kavli IPMU 
(WPI), The University of Tokyo, Kashiwa, Chiba 277-8583, Japan}

\author{Kazuya \textsc{Matsubayashi}\altaffilmark{4}}
\altaffiltext{4}{Okayama Astrophysical Observatory, 
Honjo 3037-5, Kamogata, Okayama 719-0232, Japan}

\author{Takashi \textsc{Hattori}\altaffilmark{5}}
\altaffiltext{5}{Subaru Telescope, National Astronomical Observatory of 
Japan, 
Hilo, HI 96720, USA}

\author{Atsushi \textsc{Shimono}\altaffilmark{3}}

\author{Kentaro \textsc{Aoki}\altaffilmark{5}}

\author{Yutaka \textsc{Hayano}\altaffilmark{2}}

\author{Yosuke \textsc{Minowa}\altaffilmark{5}}

\author{Kazuma \textsc{Mitsuda}\altaffilmark{6}}
\altaffiltext{6}{
Astronomical Institute, Tohoku University, 6-3 Aramaki, 
Sendai, Miyagi 980-8578, Japan}

\author{Yasuhito \textsc{Hashiba}\altaffilmark{7}}
\altaffiltext{7}{Institute of Astronomy, the University of Tokyo, 
Mitaka, Tokyo 181-0015, Japan}



\maketitle

\begin{abstract}
We report the initial result of 
an 
adaptive-optics assisted, optical 
integral-field-unit 
observation 
on IRAS~04576+0912, 
the 
nearest ($z=0.039$) 
active galactic nucleus 
with a prominent blueshift/tail in [O\,III] emission 
from a sample of such objects that we have collected from the literature.
We 
aim at 
addressing the putative quasar-mode feedback process 
with Subaru/Kyoto~3D~II$+$AO188. 
The 
optical waveband 
(6400--7500{\AA}) enables us to 
measure the gas density via the 
[S\,II] doublets, 
in contrast to earlier Near-IR studies. 
Since 
the fast [O III] outflow 
happens only 
around rapidly growing 
central black holes, 
this object 
is 
suitable for investigating 
the black hole-galaxy coevolution. 
The obtained data cube 
exhibits blue tail in the [S II] emission at many lenslets. 
By fitting the 
spectrum 
with the high 
excess flux at the [S II] blue tail, 
we find 
the fast ($\sim$860~km/s), 
dense ($>$3000/cc), 
wide-angle 
and 
offset 
outflow in central 100-pc scales.
Although 
the large opening angle and 
the high 
gas outflow-to-accretion ratio 
may favour the 
feedback 
hypothesis, 
the inferred kinetic power injection rate 
of this ionized gas outflow seems insufficient 
to influence the whole host 
galaxy.
A conventional assumption 
of a low 
density 
must have overestimated 
the feedback process.
\end{abstract}

\section{Introduction}

Various 
investigations 
on the evolution of galaxies and massive black holes (BHs)
indicate
a strong outflow, driven by the radiation from the accretion disk
of active galactic nuclei (AGNs) 
(e.g., Silk \& Rees 1998). 
It is required to {self-regulate star formation} of host galaxies and 
to hamper producing too massive galaxies 
(``quasar-mode feedback''; 
e.g., Schawinski et al. 2007).
There are many evidences for outflows from AGNs, 
observed as 
C\,IV bluetail, 
[O\,III] blueshift, 
broad 
and 
narrow absorption lines (BALs and NALs), 
X-ray 
and 
many other lines 
(e.g., Sulentic et al.\ 2007; Zamanov et al.\ 2002).
Winds from quasars may have played a major role 
in the BH-galaxy coevolution 
(e.g., 
Wyithe \& Loeb 2003; 
Di Matteo et al. 2005). 

However, 
a similarly numerous 
works 
indicate that the AGN feedback is not efficient enough 
(Gabor et al. 2014; Balmaverde et al. 2016; 
Carniani et al. 2016; Villar-Martin et al. 2016). 

Due to the lack of 
density indicators or
sufficient spatial resolution in earlier observations, 
basic questions on the putative 
quasar-mode feedback, such as 
if there is really the quasar-mode feedback, 
if it is powerful enough to quench 
star formation, or 
if every AGN launches the feedback, 
are still unclear.

We then carried out 
an {adaptive-optics (AO) assisted, optical} 
{integral-field-unit (IFU) observations for nearby AGNs 
that are plausible candidates in the 
act of feedback}.
Our targets are selected from objects 
that evidently show blueshifted components in 
[O\,III] emission 
in conventional long-slit spectroscopic data. 

{Optical} wave-band we chose (6400--7500{\AA}) enables us to {measure the 
gas density} via the flux ratio of the [S\,II]$\lambda \lambda $6716, 6731\AA \ 
doublet emission lines (Osterbrock \& Ferland 2006). 
AO-IFUs other than Kyoto\,3D\,II work only at Near-IR band, 
and hence previous AO-IFU studies assumed 
the density 
(e.g., Storchi-Bergmann et al.\ 2009; 
Mazzalay et al.\ 2013), resulting in large uncertainties in 
the mass outflow rate and the kinetic power etc of the 
feedback 
process. 

We use Kyoto\,3D\,II with AO188 
mounted on Subaru telescope 
(Sugai et al. 2010; 
Matsubayashi et al. 2016). 
Long-slit spectroscopic data 
for AGN outflows 
(e.g., broad wings of emission lines, BALs and NALs)
have uncertainties in the geometry and the covering factor, 
and thus 
to what degree the 
outflow influences 
the interstellar medium (ISM) 
is unclear. 
Previous optical IFU observations 
were not implemented with AO (e.g., Barbosa et al.\ 2009; 
Cresci et al.\ 2015; Lena et al.\ 2015). 
The {AO assistance 
is critical} for 
both mapping the outflow 
with a high spatial sampling (Husemann et al. 2016) and 
substantial reduction of the major noise source 
(i.e., broad lines and continuum emission).

In this Letter, 
we report the initial result of our 
observation 
for the nearest AGN 
with prominent [O III] blueshift 
from a sample of such objects that we have collected from the literature.
In the next section, 
we briefly describe the selection for 
our target. 
Then, observational setups and data 
reduction are shortly summarised.
In \S 4, 
we present the results 
on the 100~pc scale, fast and dense outflow.
Finally, we make a summary of this study and discussion in \S 5.
We adopt 
the standard $\Lambda$CDM cosmology, 
with $\Omega_\Lambda =0.714$, 
$\Omega_M =0.286$ and $H_0 = 69.6$\,km\,s$^{-1}$\,Mpc$^{-1}$. 

\section{Target Selection}\label{sec:target}

We regard that 
targets selected 
from 
{[O\,III] 
blueshifting 
samples are desirable 
to examine the putative AGN feedback process. 
This selection 
ensures that the outflow 
takes place in 
the narrow-line region (NLR),  
overflowing 
beyond the gravitational potential well 
of the central BH 
($\sim$1\,pc for a $10^7 M_\odot$ BH).
Outflow in 
the broad line region 
(at $\sim$0.01\,pc from the center; observed 
in 
C\,IV, X-ray or BALs) 
is not sure whether it overflows to the galactic scale ($\sim$kpc)
or it is to fall back again. 

Our targets 
are narrow-line Seyfert 1 galaxies (NLS1s),
 harbouring 
rapidly growing massive BHs (Kawaguchi 2003; Kawaguchi et al.\ 2004). 
The [O III] blueshift
(with velocity up to $\sim$\,1000\,km/s)
occurs 
only at the phase with 
high accretion rates onto central BHs 
(namely, mostly in NLS1s; 
Aoki, Kawaguchi \& Ohta 2005; Marziani et al. 2003;  
Boroson 2005; Komossa et al.\ 2008).
If the 
AGN feedback 
process is really efficient, we will then unveil 
the 
{BH-galaxy coevolution}
in 
{the early stage of 
BH growth}.
The outflow velocity depends on neither the radio power  
nor radio loudness (Aoki et al. 2005), indicating that the outflow 
is not powered by small-scale jets.

AGNs with prominent blueshifts in [O III] emission lines are 
summarised by Aoki et al.\ (2005) and Komossa et al.\ (2008).
To ensure the {highest spatial resolution in physical scale}, 
 we chose the {nearest one} 
among the objects with measured/expected NLR size larger than 1"
in 
the two literatures.
The nearest object, IRAS\,04576+0912 at redshift $z$ of 
$\sim$0.039 (\S~\ref{ssec:vel_den}), 
is a SBa galaxy (Ohta et al. 2007).
The field-of-view (FoV) 
and the spatial sampling 
of Kyoto\,3D\,II ($3.1" \times 2.4"$ 
and $0.084"$) 
corresponds to 
2.4$\times$1.9\,kpc and  
65~pc, respectively.
Its 
optical spectrum 
(V{\'e}ron-Cetty et al.\ 2001) 
shows the blueshift in [O III]$\lambda \lambda $4959, 5007{\AA}  
and [N\,II]$\lambda$6584{\AA} emission lines, 
by $\sim$300
\,km/s (Aoki et al. 2005) from the 
velocity 
that matches with 
the Balmer lines. 

The FWHMs of H$\alpha$ and H$\beta$ emission lines are 
1100~km/s and 1210--1220~km/s, respectively 
(V{\'e}ron-Cetty et al.\ 2001).
The estimated BH mass, the bolometric luminosity $L_{\rm bol}$
and the Eddington ratio are 
$10^{6.6} M_\odot$,
6.2\,$\times$\,10$^{44}$\,erg/s  
and 1.15, respectively (Aoki et al. 2005).
In a diagram of the BH mass versus the optical luminosity (Kawaguchi 2003), 
IRAS\,04576 is a typical NLS1 and its accretion rate onto the central BH is 
estimated to be $\sim$\,0.8\,$M_\odot$/yr.

\section{Observation and Data Reduction}

On the night of 24 September 2015, we observed IRAS~04576 
and the spectrophotometric standard star (BD+21~607) 
with Subaru/Kyoto\,3D\,II+AO188 in laser guide star (LGS) mode.
Unfortunately, we could use only a half of the FoV due to a 
temporal detector trouble.
We took 7 frames, with 1200\,sec exposure time for each frame.
We used the nearby star, 17" apart from the center to NE with 
$V \sim 15.4$~mag, 
as the tip-tilt guide star.
During the exposures, 
the image size (FWHM) of the guide star was about 0."35--0."45.
The spectral resolution  
is $R\sim$\,1200, and the wavelength coverage is 6400--7500~{\AA}. 

Data reduction is carried out using custom made IRAF scripts 
(Sugai et al.~2010) adapted for the newly installed deep 
depletion CCD (Mitsuda et al.~2016). 
The reduction process includes bias subtraction, 
spectrum extraction, flat fielding. 
Cosmic-rays were removed using L.A.Cosmic (van Dokkum 2001). 
After wavelength calibration, sky subtraction was performed 
referring the sky aperture spectra. 
Kyoto~3D~II simultaneously obtains the spectra of the object 
and the sky $\sim 29"$ away from the object field. 
Flux was calibrated using the standard star. 
Atmospheric absorption features were corrected using the normalized 
spectrum of the standard star because it is an early type (F2) star.
After converting 
to the heliocentric velocity,
the 7 frames are combined.
Then, spectral fitting throughout this study is carried out 
by our own python code.

\section{Line Fitting and the Results}

Before fitting 
the narrow emission lines, we 
fix the two issues that affect their measurements, 
in the next subsection.

\subsection{Laser-Induced Sky Emission and Broad Emission Line}\label{ssec:absbroad}

The LGS laser 
excites 
atmospheric molecules, 
enhancing 
their sky emissions via the Raman scattering (Vogt et al. 2017). 
The enhancement 
is weaker at the science 
FoV 
than the sky field, 
since the 
secondary mirror tends to hide the 
former from the scattered light.
As a result of sky subtraction, 
an artificial "absorption" line, at $\sim 6827$~{\AA} (between 
H$\alpha$ and [N~II]\,$\lambda$6583{\AA} lines) due to 
 N$_2$ molecules in our case, appears in each science spectrum.
We fit the sky spectrum with a Gaussian, and determined the wavelength and the width. 

A Fabry-Perot, narrowband engineering 
observation in February 2015 
showed that the 
"absorption" depth 
is constant 
for the whole science FoV. 
By fitting the spectra at 
the 
north-east and the west ends, showing the clearest "absorption" 
(with the negative flux density at the corresponding wavelengths),  
we derive the mean depth of the "absorption", 
and use it 
for all the analysis hereafter.

Next, we fix the shape of the broad H$\alpha$ emission line. 
Since the intrinsic 
size of the broad line region is about 0.01~pc
($\ll $
the angular size of 1 lenslet) 
and the apparent spatial extent is due to the 
point-spread-function (PSF) smearing, 
the line profile should be common among lenslets.
We examine 
which function, 
among single Gaussian, double Gaussian and single Lorentzian, 
can successfully fit the broad line simultaneously in 
different spatial regions (four regions in middle left panel of 
Fig.~\ref{fig:maps}).
We find 
the double Gaussian 
most successful.
Of the two components of the double Gaussian, 
the narrower and higher one has the central wavelength and the FWHM 
of 6817.96~{\AA} and 1306~km/s, while 
the broader and lower one has 6814.41~{\AA} and 2579~km/s, respectively.
The peak 
ratio between the second over the first of 
0.49   
is 
fixed for the entire FoV.

\subsection{Fit with Single Narrow-Line Component}\label{ssec:1comp}

For the 615 lenslets that have large enough flux at 
line-free wavelengths
compared with the sky spectrum, 
we fit each spectrum at 6600--7200~{\AA} by a combination of 
a linear continuum,
a broad H$\alpha $ emission line (\S~\ref{ssec:absbroad}) and  
a single Gaussian velocity component of narrow emission lines 
with a 
common line-width
composed of H$\alpha $, 
[N II]$\lambda \lambda $6548, 6583{\AA} and 
[S II]$\lambda \lambda $6716, 6731{\AA}.
The line ratio of [N II] is fixed so that the 6548{\AA}/6583{\AA} flux 
ratio is 
0.34. 
In figure \ref{fig:maps}, 
maps of various line flux, flux ratio and 
the velocity are shown.

\begin{figure*}
 \begin{center}
  \includegraphics[width=\textwidth]{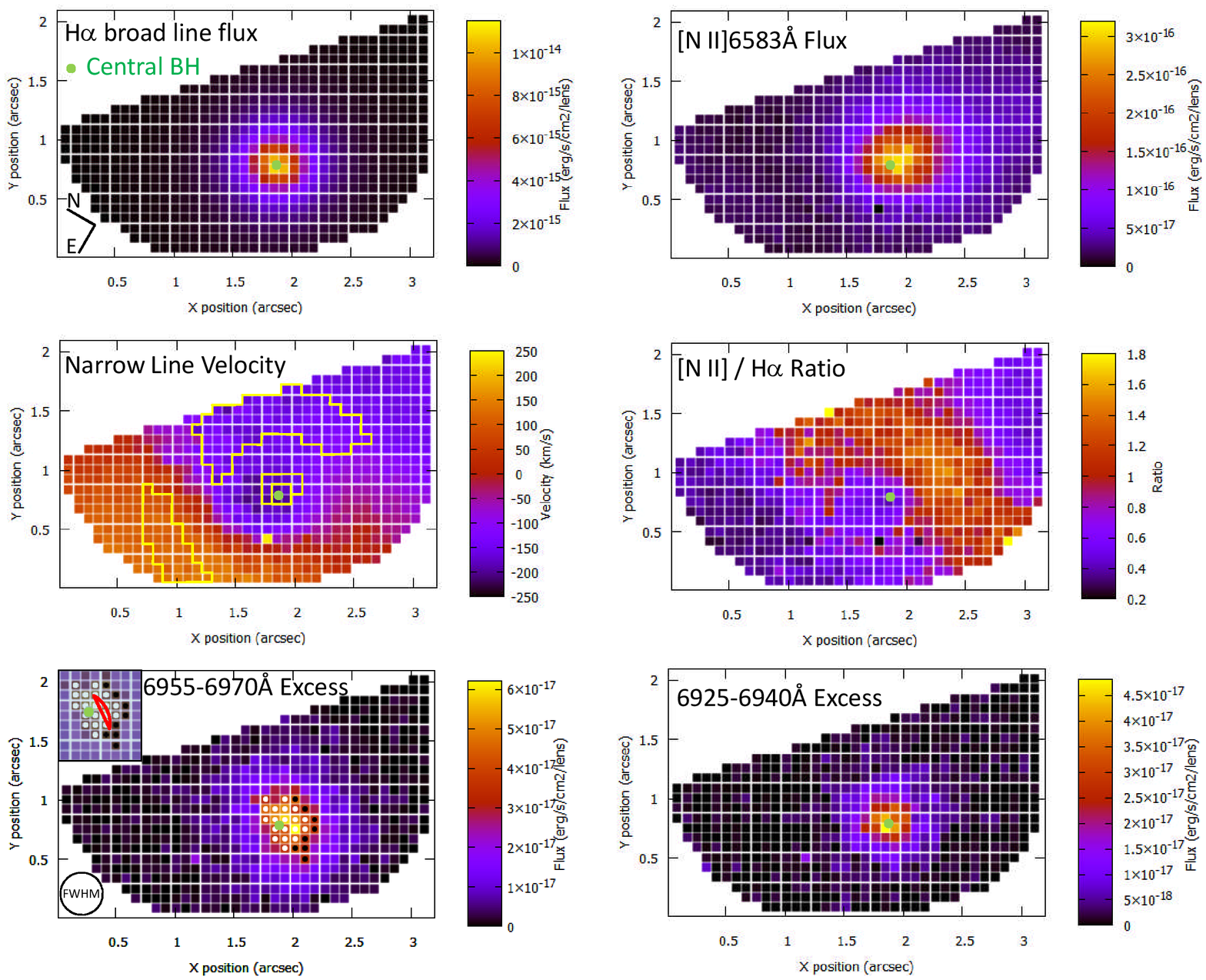} 
 \end{center}
\caption{
Maps obtained through the fitting with a single component for narrow emission lines;
H$\alpha$ broad emission flux (top left), [N II]~6583{\AA} narrow emission flux (top right), 
velocity of narrow emission lines (middle left, with yellow four regions that are 
used for determining the broad line shape; \S~\ref{ssec:absbroad}), 
the line ratio between [N II]~6583{\AA} over 
H$\alpha$ narrow emission (middle right), 
excess flux integrated over 6955--6970{\AA} (bottom left) and 
that at 6925--6940{\AA} (bottom right).
The 
horizontal range of 3."1 corresponds to 2.4~kpc.
A 
circle at the left-bottom panel 
has a diameter of the PSF FWHM, 0."37 (290~pc), and 
the inset shows a possible intrinsic structure 
for the outflowing region. 
When drawing the velocity map, a systemic velocity is 
chosen at 
$z$ of 
0.0388 (\S~\ref{ssec:vel_den}).
}
\label{fig:maps}
\end{figure*}

The 
flux map of the 
broad 
H$\alpha$ 
line 
represents 
the 
PSF during our observation, and 
turns out a round shape (top left in Fig.~\ref{fig:maps}).
By fitting the 
map, 
we obtain the central position 
and the radial profile, 
which is well described by a Moffat profile with the FWHM of 0."37. 
The PSF FWHM is similar to the image size of the guide star.
These indicate successful fitting. 
The central position indicates the position of the 
BH.

The 
[N II]$\lambda $6583{\AA} flux map (upper right)
shows an elongation toward 
the west 
from the central BH. 
Maps for the velocity of narrow emission lines (middle left)
and 
the [N II]/H$\alpha $ flux ratio 
(middle right)
are also 
asymmetric 
in kpc-scales.
The overall velocity map shows blueshifts on west 
and redshifts on the east 
sides of the BH, 
indicating a galactic rotation with an 
axis inclined to 
the north 
direction.
Alternatively, the velocity gradient may imply 
the double-sided outflowing NLR.
Careful decomposition into the galactic rotation 
and the NLR kinematics will be needed.
The [N II]/H$\alpha$ flux ratio is larger than $\sim 1.0$ at the west 
of the nucleus, strongly suggesting the AGN origin for the 
ionization (e.g., Kauffmann et al. 2003). 
Around the nucleus and $\sim$\,1" to the west, the ratio is relatively
lower, but still larger than the border value between 
AGN- and star formation-origins ($\sim$\,0.5).
At 1"--1."5 northeast of the nucleus, the ratio is even smaller 
down to $\sim$\,0.3, indicating that there is a star forming region.

\subsection{Outflowing Region}\label{ssec:out_region}

When fitting the 615 spectra, it turned out that 
many lenslets 
show 
blue tail in [S II] emission lines (figure \ref{fig:spec_fit}), 
indicating 
an outflowing component of [S II] emission.

\begin{figure*}
 \begin{center}
  \includegraphics[width=0.75\textwidth]{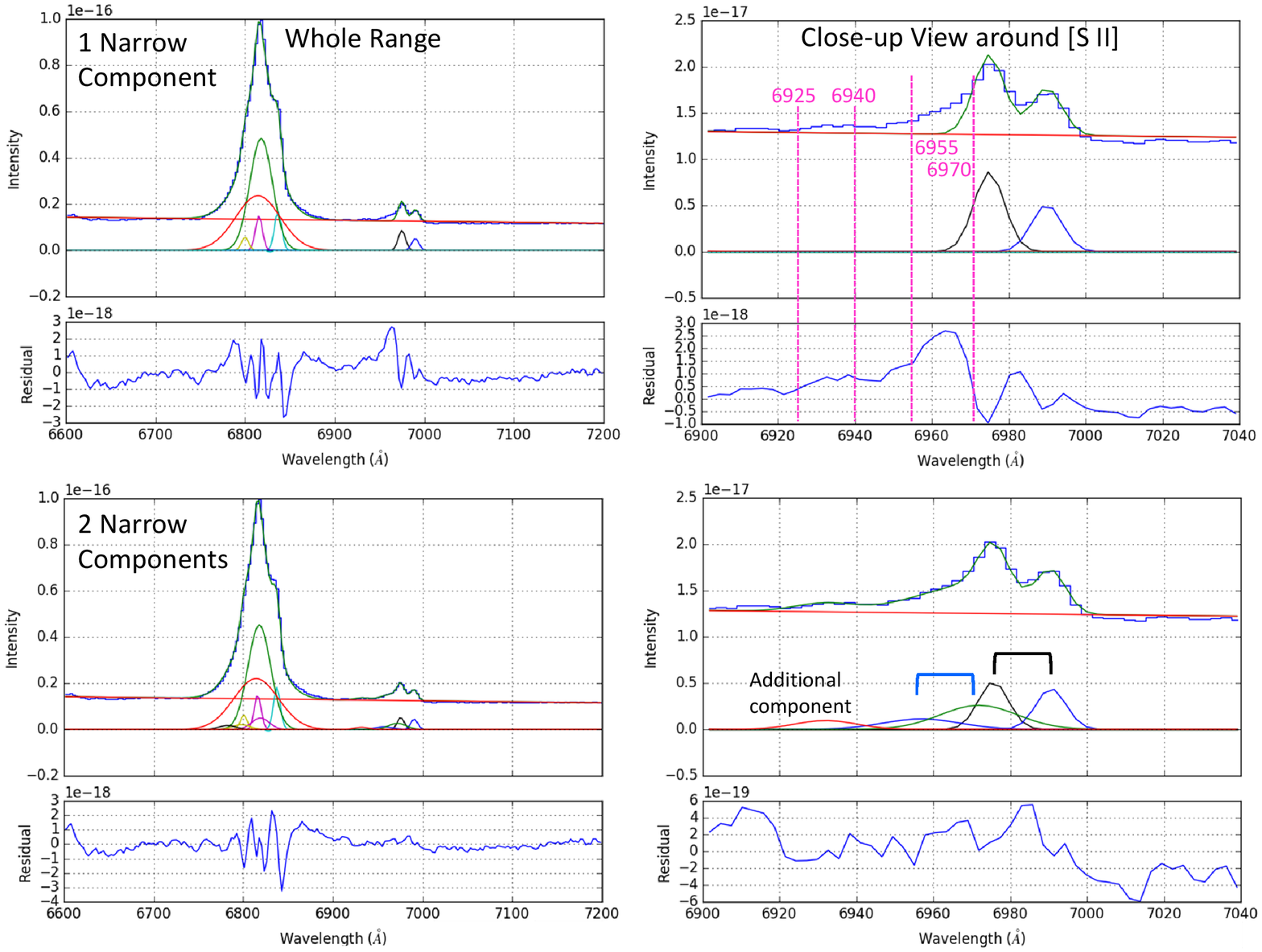} 
 \end{center}
\caption{
Spectral fitting for the mean spectrum of the large 6955--6970{\AA} excess 
region.
Left panels exhibit the whole wavelength range, 
whereas 
right panels show the close-up views around the [S II] emission.
Upper panels present the result with a single narrow component, while 
lower panels are obtained with two narrow components.
Vertical dotted lines indicate the two wavelength ranges 
for the excess flux maps (bottom panels 
in Fig.~\ref{fig:maps}). 
}
\label{fig:spec_fit}
\end{figure*}

To identify 
the outflowing region, we calculate the 
excess flux of the observed data over the model 
(\S~\ref{ssec:1comp}) 
at 6955--6970{\AA}. 
Bottom-left panel in figure \ref{fig:maps} shows the excess flux map, 
where 
we overdraw 27 (white and black) circles 
if the excess flux is larger than the half of the maximum excess flux
(i.e., larger than $3.1 \times 10^{-17}$\,erg/s/cm$^{-2}$/lens).
Black circles mean the 8 lenslets with 
the broad line flux 
smaller than a certain threshold. 
We found that the distribution of the 
white and black circles are not 
located symmetrically around 
the central BH: 
it is mainly distributed at the 
west 
of the 
BH.
The fact that the outflowing region is offset from the center 
implies that we are not observing the outflow from the 
pole-on view.

Moreover, the area with the excess flux larger than the half of 
its peak looks elongated, and 
the elongation is nearly 
perpendicular to the radial direction from the center.
The width is comparable to the PSF FWHM, while the 
length is about 1.4 times the FWHM.
Taking account of the image smearing by the PSF, 
the distribution 
indicates that 
the outflowing 
region is intrinsically 
narrow 
(with a length of 
$\sqrt{1.4^2-1^2} \approx 1$ times the FWHM; the inset of 
bottom-left panel).
The excess flux region offsets from the BH by 
 $\sim$0."15 ($\sim$120~pc projected distance), 
implying 
a short-duration (rather than 
a continuous) outflow. 
From the estimated length 
(1$\times$FWHM) and the offset, 
we estimate the half-opening angle to be 
$50^\circ$ [$\arctan \{(0."37/2)/0."15 \}$].
The inferred angle indicates that the outflow is not 
jet-like (nor spherical).
Since 
a narrow jet-like geometry 
would have difficulty to push out 
or disturb the ISM of its host galaxy in a large volume, 
the wide-angle estimated above is likely 
in favour of the AGN feedback hypothesis. 

There is a small bump in the spectrum at around 6930\,{\AA} (Fig.~\ref{fig:spec_fit}).
To 
investigate its 
origin,
we show the excess flux map at  6925--6940{\AA} as well (bottom right panel).
The distribution turns out 
similar to that of the broad H$\alpha $ 
emission. 
Namely, this spectral feature is not associated with the region 
responsible for the [S II] blue tail. 
Instead, this bump seems 
due to either 
[S II] line(s) emitted by another outflowing component 
with a higher velocity 
near the BH, 
highly redshifting H$\alpha $ emission from a blob in the broad line region, 
or something else (e.g., He\,I$\lambda 6678${\AA}).
The flux of this bump is consistent with the Case B, 
He\,I$\lambda 6678${\AA} recombination at the broad-line region 
having the solar abundance, the temperature of $10^4$\,K 
and the electron density of $10^4$\,cm$^{-3}$ (Osterbrock \& Ferland 2006), 
if all (or almost all) the \atom{He}{}{} are \atom{He}{}{}$^+$ ions.

Solving its origin is beyond the scope of this study. 
This component, together with the 
broad H$\alpha $ wing, 
could influence 
measuring 
the outflowing gas. 
In the next subsection,
we thus preferentially use the spectra at the lenslets far from the center
(indicated by black circles in the bottom left panel in Fig.~\ref{fig:maps}).

\subsection{Velocity and Density of the Outflowing Gas}\label{ssec:vel_den}

To derive the velocity and the density of the outflowing gas, 
we use the mean spectrum of the 8 lenslets where the excess 
flux is large and the contamination of the broad line is small. 
If we fit the 
spectrum with a single velocity component 
for narrow lines, the clear blue tail in [S II] emission 
remains 
(upper panels in Fig.~\ref{fig:spec_fit}).
Moreover, the flux ratio of [S II] emission 
lines is 
1.7, 
exceeding the allowed range 
(0.44--1.45; e.g., Sanders et al. 2016).
These issues 
indicate that the fitting is not appropriate.

Therefore, we fit the mean spectrum with 
additional velocity 
component for narrow emission lines (H$\alpha$, [N~II] and [S~II]).
Similar to the preexisting narrow line component, 
the width and the velocity 
are tied too within this second component.
Here, we also add a component at $\sim$6930{\AA}, for which 
the wavelength and the gaussian width are fixed at the lenslet 
with the strongest 6925--6940{\AA} excess flux (shown by yellow in 
bottom right panel in Fig.~\ref{fig:maps}).
As the result (Table~{\ref{tab:2comp}}), 
the blue tail is well described by the 
second 
narrow 
component 
(lower panels in Fig.~\ref{fig:spec_fit}).
This component is about 810\,km/s 
blueshifted from the brighter component at these lenslets, 
which
is relatively blueshifted with respect to 
the whole FoV 
(middle left panel of Fig.~\ref{fig:maps}).
Since the systemic velocity 
of 
this object 
in literature varies by hundreds km/s, 
we here define 
the systemic velocity as 
the mean velocity of the narrow lines at the north-east and 
west sides $\gtrsim$\,1" apart from the center.
The blueshift velocity of the outflowing component relative to 
the systemic 
velocity 
is about 860\,km/s, 
fast enough compared with 
the escape velocity at NLRs ($\sim$500km/s). 

If the gas density is low ($\lesssim 100$/cc), 
the [S~II] flux ratio (6717{\AA}/6731{\AA}) 
is about 1.4--1.45.
When the density exceeds 
$\sim$~$10^{3.5}$/cc, 
the [S~II] flux ratio becomes 
$\sim$0.44--0.5 (e.g., Sanders et al. 2016).
By the 2-component fit, 
we estimate the 
brighter component has the density of about 300/cc, similar 
to a typical density in NLRs of AGNs.
On the other hand,
the outflowing component is very dense, 
$>$~3000/cc. 


\begin{table}
  \tbl{Fit with Two Narrow 
Components.}{%
  \begin{tabular}{ll}
      \hline
\multicolumn{2}{c}{Brighter Component} \\ 
      \hline
[S~II]$\lambda $6731 central wavelength [{\AA}] & 6990.63$\pm$0.10 \\
FWHM [km/s] & 392.3$\pm$14.4  \\
Flux density (6716~{\AA})$^a$ & 5.16$\pm$0.82  \\
Flux density (6731~{\AA})$^a$ & 4.43$\pm$0.60  \\
Flux ratio (6716/6731) & 1.17$\pm$0.24  \\
Gas density [cm$^{-3}$] & $\sim$300 (20--800)  \\
      \hline
      \hline
\multicolumn{2}{c}{Blueshifted Component} \\ 
      \hline
[S~II]$\lambda $6731 central wavelength [{\AA}] & 6971.9$\pm$1.4 \\
FWHM [km/s] & 1004$\pm$168  \\
Flux density (6716~{\AA})$^a$ & 1.11$\pm$0.48  \\
Flux density (6731~{\AA})$^a$ & 2.60$\pm$0.64  \\
Flux ratio (6716/6731) & 0.43$\pm$0.21  \\
Gas density [cm$^{-3}$] & $>$~3000  \\
Narrow H$\alpha$ Flux [erg/s/cm$^{2}$/lens] & (5.0$\pm$1.5)$\times 10^{-17}$  \\
      \hline
    \end{tabular}}\label{tab:2comp}
\begin{tabnote}
$^a$ Peak intensity in the unit of $10^{-18}$~[erg/s/cm$^2$/{\AA}/lens].
\end{tabnote}
\end{table}


Albeit large uncertainties,
the [N~II]$\lambda $6583/H$\alpha$ 
and [S~II]$\lambda $6716+$\lambda $6731/H$\alpha$
flux ratios of the outflowing
component ($2.4 \pm 0.8$ and $1.8 \pm 0.6$, respectively) 
indicate the non-stellar origin for its ionization (e.g., Kewley et al. 2006).

Assuming Case B recombination at 
10$^4$~K with the solar abundance,  
the ionized gas mass $M_{\rm gas}$ is in principle estimated from the 
H$\alpha$ luminosity $L_{H\alpha}$
and 
the electron density $n$, 
as 
$M_{\rm gas} \approx 3.3 
\times 10^8 M_\odot \, 
(L_{H\alpha}/10^{43}\,[{\rm erg/s}]) \, (n/100 [{\rm cm}^{-3}])^{-1}$ 
(Carniani et al. 2015; Nesvadba et al. 2017).
Since 
the gas density affects estimating 
the gas mass 
inversely, 
a conventional assumption for the density 
of $\sim$\,100/cc 
must have led overestimations in the gas mass, the outflow rate, 
the kinetic power injection etc for the feedback process. 
The flux of H$\alpha $ narrow line of the outflowing component 
is quite uncertain, since it is veiled by the strong broad line.
Taking the fitting results at face value, 
and assuming that 
the 8 lenslets contain the 10\% 
of the 
total (all the 27 lenslets plus surrounding lenslets leaked 
via the image smearing) flux of H$\alpha $ narrow line in
the outflowing gas, 
we estimate the outflowing gas mass 
to be $< 1.6  
\times 10^4\,M_\odot$. 
Given that the radial extension of the outflowing region is 
much smaller than the PSF FWHM (0."37), 
we presume 
that it is 0.1$\times$0."37 (i.e., 29\,pc).
Assuming that the line-of-sight velocity (860~km/s) 
is $\cos 45^\circ$ (\S~\ref{ssec:out_region}) times 
the outflow 
velocity $V_{\rm out}$ (i.e., 1200\,km/s), 
this region is likely launched from the center 0.13\,Myr ago 
for 
a duration of 2.3$\times$10$^4$\,yr (i.e., 29pc/1200km/s; $t_{\rm out}$). 
Then, 
the gas outflow rate,  $ M_{\rm gas}  / t_{\rm out} $, 
is estimated to be 
$< 0.7 \, M_\odot$/yr ($\sim$\,90\% of 
the current gas accretion rate; \S~\ref{sec:target}).

Similarly, 
the kinetic power $M_{\rm out} (V_{\rm out}^2 + 3 \sigma^2)/ 2$ 
and the momentum $M_{\rm out} V_{\rm out}$
contained in the outflowing gas lead to estimations for 
the kinetic energy injection rate 
($< 4.4 \times 10^{41}$\,erg/s) 
and 
the momentum flux 
($< 5.3 \times 10^{33}$\,g\,cm/s), where 
$\sigma$ is the 
line-of-sight velocity dispersion 
of the blueshifted 
line.
The ratio of the kinetic energy injection rate over the 
bolometric luminosity is low 
($\lesssim$\,0.07\%), 
similar to 
high-$z$ AGNs 
(Kakkad et al. 2016).
The momentum flux is 
comparable to 
the radiation momentum $L_{\rm bol}/c$ received by the 
outflowing region with a solid angle subtended at the nucleus 
of $2 \pi (1-\cos 50^\circ)$ ($=$\,0.18\,$\times$4$\pi$), 
where $c$ is the speed of light.

\section{Summary and Discussions}

We observe 
the 
nearest ($z=0.039$) AGN 
with prominent blueshift/tail in [O\,III] emission
from a sample of such objects that we have collected from the literature, 
to examine 
the putative AGN feedback process on its host galaxy, 
by utilising the unique capability of Subaru, 
optical IFU with AO assistance (Kyoto~3D~II with AO188).
The optical waveband enables us to 
observe 
the density-sensitive [S II] doublets, 
with which we can measure the gas density via the flux ratio.
Through the spectral fitting for the 615 lenslets, over 
3."1 (2.4\,kpc) 
of IRAS~04576+0912, 
we obtain the kpc-scale structures 
in velocity, flux and ionization sources.
Many lenslets turn out to exhibit blue tail in the [S II] emission.
Through the location and the distribution of the lenslets 
with the high 
excess flux at the corresponding wavelengths, 
and fitting results for their mean spectrum, 
we found 
the fast ($\sim$860~km/s blueshift), 
dense ($>$~3000/cc), 
wide-angle (with the half-opening angle of $\sim 50^\circ$) and 
offset outflow in 100-pc scales.
In short, a gas outflow with a rate of 
$\lesssim$\,90\% of the accretion rate was launched 0.13\,Myr ago 
(over 2.3$\times$10$^4$\,yr). 
The large opening angle and 
the high gas outflow-to-accretion ratio 
may favour the AGN feedback 
hypothesis, 
although the outflow seems not to be spherical nor continuous.
The inferred kinetic power injection rate 
($\lesssim$\,0.07\% of $L_{\rm bol}$) 
of this ionized gas outflow seems insufficient 
to influence the whole host 
galaxy.
A conventional assumption 
of a low gas density, 
$\sim$\,100/cc, 
must have overestimated  
the feedback process. 

Outflow launched by AGNs with high accretion rates 
may be commonly fast and dense.
AO-assisted NIR IFU observations on 
NGC~1068, which 
is a type-2 analogue of NLS1s (Kawaguchi 2003), 
reveal the outflow kinematics 
in the central 100pc-scales 
(M{\"u}ller-S{\'a}nchez et al. 2011). 
Ozaki (2009) showed that the outflowing clumps 
(with gas densities of $\sim$10$^5$/cc) with a variety of 
column densities are likely accelerated by radiation from the center. 

We leave 
further analysis, 
including maps 
for other line ratios 
and 
the narrow-line width (as a tracer for ISM disturbance)
as well as the decomposition between the 
galactic rotation and the NLR outflow, 
for a next paper.
Three-dimensional kinematical modelling to reproduce these 
maps will follow.

\begin{ack}
This work is based on data collected 
at Subaru Telescope, which is operated by the 
National Astronomical Observatory of Japan.  
We are grateful to the staff of the Subaru Telescope,
and 
the anonymous referee,  
Masayuki Akiyama and Xiaoyang Chen for helpful comments.
TK is supported 
by 
JSPS KAKENHI (17K05389).
\end{ack}



\begin{thebibliography}{}

\bibitem[Aoki et al.(2005)]{2005ApJ...618..601A} Aoki, K., Kawaguchi, T., \& Ohta, K.\ 2005, \apj, 618, 601 

\bibitem[Balmaverde et al.(2016)]{2016A&A...585A.148B} 
Balmaverde, B., Marconi, A., Brusa, M., et al.\ 2016, \aap, 585, A148 

\bibitem[Barbosa et al.(2009)]{2009MNRAS.396....2B} 
Barbosa, F.~K.~B., Storchi-Bergmann, T., Cid Fernandes, R., 
Winge, C., \& Schmitt, H.\ 2009, \mnras, 396, 2 


\bibitem[Boroson(2005)]{2005AJ....130..381B} Boroson, T.\ 2005, \aj, 130, 381 

\bibitem[Carniani et al.(2015)]{2015A&A...580A.102C} 
Carniani, S., Marconi, A., Maiolino, R., et al.\ 2015, \aap, 580, A102 

\bibitem[Carniani et al.(2016)]{2016A&A...591A..28C} 
Carniani, S., Marconi, A., Maiolino, R., et al.\ 2016, \aap, 591, A28 


\bibitem[Cresci et al.(2015)]{2015A&A...582A..63C} Cresci, G., Marconi, A., Zibetti, S., et al.\ 2015, \aap, 582, A63 

%

\bibitem[Di Matteo et al.(2005)]{2005Natur.433..604D} 
Di Matteo, T., Springel, V., \& Hernquist, L.\ 2005, \nat, 433, 604 



%

\bibitem[Gabor \& Bournaud(2014)]{2014MNRAS.441.1615G} 
Gabor, J.~M., \& Bournaud, F.\ 2014, \mnras, 441, 1615 

\bibitem[Husemann et al.(2016)]{2016A&A...594A..44H} 
Husemann, B., Scharw{\"a}chter, J., Bennert, V.~N., et al.\ 2016, \aap, 594, A44 

\bibitem[Kakkad et al.(2016)]{2016A&A...592A.148K} 
Kakkad, D., Mainieri, V., Padovani, P., et al.\ 2016, \aap, 592, A148 

\bibitem[Kauffmann et al.(2003)]{2003MNRAS.346.1055K} 
Kauffmann, G., Heckman, T.~M., Tremonti, C., et al.\ 2003, \mnras, 346, 1055 

\bibitem[Kawaguchi(2003)]{2003ApJ...593...69K} Kawaguchi, T.\ 2003, \apj, 593, 69 

\bibitem[Kawaguchi et al.(2004)]{2004A&A...420L..23K} 
Kawaguchi, T., Aoki, K., Ohta, K., \& Collin, S.\ 2004, \aap, 420, L23 

\bibitem[Kewley et al.(2006)]{2006MNRAS.372..961K} 
Kewley, L.~J., Groves, B., Kauffmann, G., \& Heckman, T.\ 2006, \mnras, 372, 961 

\bibitem[Komossa et al.(2008)]{2008ApJ...680..926K} 
Komossa, S., Xu, D., Zhou, H., Storchi-Bergmann, T., \& Binette, L.\ 2008, \apj, 680, 926-938 


\bibitem[Lena et al.(2015)]{2015ApJ...806...84L} 
Lena, D., Robinson, A., Storchi-Bergman, T., et al.\ 2015, \apj, 806, 84 

\bibitem[Marziani et al.(2003)]{2003MNRAS.345.1133M} 
Marziani, P., Zamanov, R.~K., Sulentic, J.~W., \& Calvani, M.\ 2003, \mnras, 345, 1133 

\bibitem[Matsubayashi et al.(2016)]{2016PASP..128i5003M} 
Matsubayashi, K., Sugai, H., Shimono, A., et al.\ 2016, \pasp, 128, 095003 


\bibitem[Mazzalay et al.(2013)]{2013MNRAS.430.2411M} 
Mazzalay, X., Rodr{\'{\i}}guez-Ardila, A., Komossa, S., \& McGregor, P.~J.\ 2013, \mnras, 430, 2411 

\bibitem[Mitsuda et al.(2016)]{2016SPIE.9908E..2MM} 
Mitsuda, K., Hashiba, Y., Minowa, Y., et al.\ 2016, \procspie, 9908, 99082M 


\bibitem[M{\"u}ller-S{\'a}nchez et al.(2011)]{2011ApJ...739...69M} 
M{\"u}ller-S{\'a}nchez, F., Prieto, M.~A., Hicks, E.~K.~S., et al.\ 2011, \apj, 739, 69 

\bibitem[Nesvadba et al.(2017)]{2017A&A...599A.123N} 
Nesvadba, N.~P.~H., De Breuck, C., Lehnert, M.~D., 
Best, P.~N., \& Collet, C.\ 2017, \aap, 599, A123 

\bibitem[Osterbrock \& Ferland(2006)]{2006agna.book.....O} 
Osterbrock, D.~E., \& Ferland, G.~J.\ 2006, 
Astrophysics of gaseous nebulae and active galactic nuclei, 
2nd.~ed.
University Science Books 

\bibitem[Ozaki(2009)]{2009PASJ...61..259O} Ozaki, S.\ 2009, \pasj, 61, 259 



\bibitem[Sanders et al.(2016)]{2016ApJ...816...23S} 
Sanders, R.~L., Shapley, A.~E., Kriek, M., et al.\ 2016, \apj, 816, 23 



\bibitem[Schawinski et al.(2007)]{2007MNRAS.382.1415S} 
Schawinski, K., Thomas, D., Sarzi, M., et al.\ 2007, \mnras, 382, 1415 


%
\bibitem[Silk 
\& Rees(1998)]{1998A&A...331L...1S} Silk, J., \& Rees, M.~J.\ 1998, \aap, 331, L1 

\bibitem[Storchi-Bergmann et al.(2009)]{2009MNRAS.394.1148S} 
Storchi-Bergmann, T., McGregor, P.~J., Riffel, R.~A., et al.\ 2009, \mnras, 394, 1148 

\bibitem[Sugai et al.(2010)]{2010PASP..122..103S} 
Sugai, H., Hattori, T., Kawai, A., et al.\ 2010, \pasp, 122, 103 

%
\bibitem[Sulentic et al.(2007)]{2007ApJ...666..757S} Sulentic, J.~W., 
Bachev, R., Marziani, P., Negrete, C.~A., 
\& Dultzin, D.\ 2007, \apj, 666, 757 

\bibitem[van Dokkum(2001)]{2001PASP..113.1420V} 
van Dokkum, P.~G.\ 2001, \pasp, 113, 1420 


\bibitem[V{\'e}ron-Cetty et al.(2001)]{2001A&A...372..730V} 
V{\'e}ron-Cetty, M.-P., V{\'e}ron, P., \& Gon{\c c}alves, A.~C.\ 2001, \aap, 372, 730 

\bibitem[Villar-Mart{\'{\i}}n et al.(2016)]{2016MNRAS.460..130V} 
Villar-Mart{\'{\i}}n, M., Arribas, S., Emonts, B., et al.\ 2016, \mnras, 460, 130 

\bibitem[Vogt et al.(2017)]{2017PhRvX...7b1044V} 
Vogt, F.~P.~A., Bonaccini Calia, D., Hackenberg, W., et al.\ 2017, Physical Review X, 7, 021044 



%
\bibitem[Wyithe 
\& Loeb(2003)]{2003ApJ...595..614W} Wyithe, J.~S.~B., \& Loeb, A.\ 2003, \apj, 595, 614 



\bibitem[Zamanov et al.(2002)]{2002ApJ...576L...9Z} 
Zamanov, R., Marziani, P., Sulentic, J.~W., et al.\ 2002, \apjl, 576, L9 



\end{thebibliography}
\end{document}